\documentclass[prl,twocolumn,superscriptaddress,showpacs,epsf]{revtex4}

\usepackage{graphicx}
\usepackage{latexsym}
\usepackage{amssymb}
\usepackage{amsfonts}
\usepackage{soul}
\usepackage{color}
\begin{document}
\bibliographystyle{apsrev}

\title{Magnetic confinement of the superconducting condensate \\ in superconductor/ferromagnet hybrid composites}

\author{W. Gillijns}
\email{Werner.Gillijns@fys.kuleuven.be} \affiliation{INPAC --
Institute for Nanoscale Physics and Chemistry, Nanoscale
Superconductivity and Magnetism \\ and Pulsed Fields Group,
K.U.Leuven, Celestijnenlaan 200D, B--3001 Leuven, Belgium}
\author{A.~Yu. Aladyshkin}
\affiliation{INPAC -- Institute for Nanoscale Physics and
Chemistry, Nanoscale Superconductivity and Magnetism \\ and Pulsed
Fields Group, K.U.Leuven, Celestijnenlaan 200D, B--3001 Leuven,
Belgium} \affiliation{Institute for Physics of Microstructures,
Russian Academy of Sciences, 603950 Nizhny Novgorod, GSP-105,
Russia}
\author{A.~V. Silhanek}
\affiliation{INPAC -- Institute for Nanoscale Physics and
Chemistry, Nanoscale Superconductivity and Magnetism \\ and Pulsed
Fields Group, K.U.Leuven, Celestijnenlaan 200D, B--3001 Leuven,
Belgium}
\author{V.~V.~Moshchalkov}
\affiliation{INPAC -- Institute for Nanoscale Physics and
Chemistry, Nanoscale Superconductivity and Magnetism \\ and Pulsed
Fields Group, K.U.Leuven, Celestijnenlaan 200D, B--3001 Leuven,
Belgium}


\date{\today}
\begin{abstract}
The influence of an inhomogeneous magnetic field on the
magnetoresistance of thin Al films, used in different
superconductor/ferromagnet hybrids, has been investigated. Two contrasting magnetic textures with out-of-plane magnetization are explored, namely (i) a plain film in a multidomain state and (ii) an array of micro-sized dots. The stray fields of the  ferromagnetic structures confine the superconducting condensate  and, accordingly, modify the condition for the nucleation of  superconductivity. By switching between different magnetic states of the ferromagnet, this confinement can be tuned at will, hereby reversibly changing the dependence of the critical temperature  $T_{c}$ on an external magnetic field $H$. In particular, the  continuous evolution from a conventional linear $T_c(H)$ dependence with a single maximum to a reentrant superconducting phase boundary with multiple $T_c$ peaks has been demonstrated.
\end{abstract}

\pacs{74.78.-w 74.78.Fk 74.25.Dw}

\maketitle

The localization of a particle in a restricted volume is known to
lead to a discrete energy spectrum due to the particle's wave
nature. In some cases the trapping potential can be created
artificially in a controlled way (e.g., in quantum dots and wells), and the geometry-dependent structure of the energy levels provides a convenient way to control the optical and transport properties of such objects \cite{QuantumWells}. Remarkably, some basic concepts of standard quantum mechanics (including the confinement of the wave function) can also be applied to more complicated systems, like superconductors, in which a quantum condensate consisting of paired electrons develops. In this case, the energy of the lowest Landau level $E_{LLL}(H)$, where $H$ is the external magnetic field, determines the critical temperature $T_{c}(H)$ at which superconductivity nucleates \cite{Tinkham}. Due to the strong dependence of $E_{LLL}(H)$ on the imposed confinement, $T_{c}(H)$ for superconducting micro- and nanostructures, differs significantly from that observed in bulk superconductors \cite{Mesoscopics}. Unfortunately, once this ''geometrical'' constraint is created, the trapping potential cannot be modified any longer.

However, the use of superconductor/ferromagnet (S/F) hybrids provides an appealing alternative to localize superconducting Cooper-pairs. In such S/F hybrids the proximity effect \cite{Buzdin-2005} as well as the stray fields of the ferromagnet \cite{Lyuksyutov-2005} play an important role in changing the superconducting properties. A magnetic template which creates a nonuniform magnetic field distribution is able to localize the superconducting condensate (or normal electrons \cite{Peeters}). Such a modulated field profile can result in exotic shapes of the $T_{c}(H)$ phase boundary for superconductor/ferromagnet (S/F) hybrids, revealing a simple shift of the $T_c$ maximum towards a certain magnetic field (so called   field-induced-superconductivity \cite{Lange-03, Gillijns-06}), or a more complicated  non-monotonic $T_c(H)$ dependence with two maxima (reentrant superconductivity \cite{Aladyshkin-03, Yang-04, Gillijns-05}), and are commonly explained in terms of magnetic field compensation effects. Indeed, for thin superconducting films, placed in a nonuniform magnetic field, superconductivity first nucleates near the $|B_z|$ minima, where $B_z$ is the out-of-plane component of the total magnetic field \cite{Aladyshkin-03}. The role of the nonuniform fields is simply to locally compensate an applied magnetic field, hereby enhancing superconductivity in the compensated area and consequently obtaining a maximum $T_{c}$ at some nonzero applied field. However, not only the amplitude of the stray field, induced by the magnetic template, is of importance \cite{Gillijns-05} (as it follows from the idea of field compensation). According to the quantum size effect mentioned above, also the length scales of the area, where the compensation takes place, are crucial for the appearance of superconductivity. More precisely, localizing the superconducting order parameter (OP) in a wide region can result in a higher $T_{c}$ than a localized OP in a narrower region.

This Letter is aimed to demonstrate how {\it tunable} magnetic
confinement of the superconducting order parameter can
practically be realized. We show that this confinement is strongly dependent on the detailed structure of the underlying magnetic template. In addition, a reversible evolution of the $T_{c}(H)$ phase boundary can be obtained by changing the magnetic state of the template. These results bridge the gap between two apparently different subjects: domain-wall superconductivity and field induced superconductivity.

In order to investigate the effects of the OP
localization experimentally, two S/F hybrid samples with different ferromagnetic subsystems were investigated: a plain ferromagnetic film, containing bubble domains, and a square array of 1.52~$\mu$m sized magnetic dots with a period of 2~$\mu$m. In both cases the ferromagnets consist of a Pt(2.5 nm) buffer layer covered by a multilayer of \mbox{[Co(0.4 nm)/Pt(1.0 nm)]$_{n}$}, where $n=15$ for the plain film and $n=10$ for the dots. The resulting magnetic structures show well-defined out-of-plane magnetization \cite{Zeper-89}. Both templates are covered by a 5 nm thick Si layer followed by a superconducting Al layer of 50 nm thickness. Since the Al film is insulated from the ferromagnetic substrate, the interaction between ferromagnet and superconductor is electromagnetic in origin with negligible proximity effects. Note that due to the low upper critical field of Al, the nonuniform magnetic fields should have a stronger influence on the superconducting properties of an Al film in comparison with Pb or Nb. 

\begin{figure}[t!]
\begin{center}
\end{center}
\caption{ (color online) (a) Magnetization loops $M(H)$ of the
Co/Pt plain film at 300 K ($\vartriangle$) and 5 K ($\Box$). The magnetic field axis is normalized by the corresponding coercive field $H_{c}^{5K} = 397$ mT and $H_{c}^{300K} = 191$ mT; (b) Remanent magnetization $M_{rem}$, measured at 5 K and $H=0$  after saturation and subsequent application of a returning field $H_{ret}$ [this procedure is shown schematically in panel (c)]; (d--g) MFM pictures (5 $\times$ 5 $\mu$m$^2$) obtained at 300 K for $H_{ret}/H_{c}=$ -0.92, -1.05, -1.31 and 1.57, respectively. The dark (bright) color represents domains with positive (negative) magnetization.} \label{Fig-Magnetism}
\end{figure}

The magnetic properties of the plain Co/Pt multilayer were
investigated using a commercial Quantum Design SQUID magnetometer. Figure ~\ref{Fig-Magnetism}(a) shows the hysteresis loop at 5 K and 300 K after renormalization by their respective coercive fields $H_{c}^{5K} = 397$~mT and $H_{c}^{300K}=191$~mT.
Clearly the magnetization changes drastically for applied fields of the order of the coercive field $H_{c}$ [dark grey area in Fig. \ref{Fig-Magnetism}(a)]. This fact allows us to control the magnetization $M$ in zero externally applied field $H=0$. Indeed in Fig. \ref{Fig-Magnetism}(b) this remanent magnetization $M(H=0)$ is shown after saturating the film, applying a certain returning field $H_{ret}$ and returning back to zero field [see Fig. \ref{Fig-Magnetism}(c)] for different $H_{ret}$ values. Clearly any remanent magnetization between positive and negative saturation can be obtained by varying $H_{ret}$. To investigate the microscopic domain distribution corresponding to these remanent magnetization states, Magnetic Force Microscopy images were taken at room temperature. In Fig. \ref{Fig-Magnetism}(d--g) MFM images are shown for a selected set of returning fields, giving details about the evolution from positive to negative magnetization. Having both hysteresis loops coincide nearly perfectly indicates that similar magnetization reversal processes occur at low temperatures. Accordingly, the domain distribution is expected to undergo a similar evolution at low temperatures as well. Thus by choosing the appropriate $H_{ret}$ value the desired domain distribution can be readily prepared~\cite{Lange-02}. To control the magnetic state of the dots the same procedure can be applied since the diameter of the dots exceeds the typical size of the domains. Accordingly they are in a multidomain state \cite{Lange-03} and any intermediate remanent magnetization can be reached. 

For a better understanding of the superconducting properties in the presence of an inhomogeneous magnetic profile, the $T_{c}(H)$ phase boundaries are calculated within Ginzburg-Landau theory and are compared with the experiment. As a simplest model we assume an infinitely thin superconducting film placed on top of a periodical one dimensional domain structure [Fig.~\ref{Fig-Theory-DWS}(a) and (b)]. We account for the controllable domain distribution by changing the ratio $\alpha$ of positively $L^{(+)}$ and negatively $L^{(-)}$ magnetized domains while keeping the period $L^{(+)}+L^{(-)}$ constant. Although this relative weight $\alpha$ can be changed through the variation of the returning field $H_{ret}$, we assume $\alpha$ to be constant when measuring the superconducting properties (for more details on the model see Ref.~\cite{Aladyshkin-06}). The constancy of $\alpha$ is justified by the fact that the applied fields for measuring the superconductor are much smaller than the coercive field of the magnetic structures [see light and dark grey regions in Fig. \ref{Fig-Magnetism}(a)]. The calculated $T_{c}(H)$ phase boundaries are shown in  Fig. \ref{Fig-Theory-DWS}(c) as a function of $\alpha$. The experimental phase boundaries are extracted from magnetoresistance measurements at different temperatures [see inset Fig. \ref{Fig-Theory-DWS}(d)], using a 80\% criterium of normal state resistance and are displayed in Fig. \ref{Fig-Theory-DWS}(d).

\begin{figure}[t!]
\caption{(color online) (a) Schematic presentation of an S/F
bilayer with a 1D domain structure with different widths of
positive $L^{(+)}$ and negative $L^{(-)}$ domains; (b) Profile of the $z-$component of the magnetic
field, calculated for $L^{(-)}/L^{(+)}=0.5$ at a height of
\mbox{$h=30$~nm} using a saturation magnetization of
\mbox{$M_0=3.5\cdot 10^5$} A/m and a ferromagnetic film thickness
\mbox{$D_F=23$~nm}; (c) Field dependence of the critical
temperature $T_c$ of the considered S/F bilayer, calculated for
several ratios of $\alpha=L^{(-)}/L^{(+)}$ assuming the period
$L^{(+)}+L^{(-)}=500~$nm to be constant. The parameters of the Al film are chosen close to the experimentally determined values: critical temperature in zero field $T_{c0}=1.4$~K and coherence length $\xi(0)=100$~nm. (d) The experimental phase boundaries $T_c(H)$ for the bilayered sample in various magnetic states, obtained through different returning fields: $H_{ret}/H_{c}=0$~(I), $-1.05$~(II), $-1.11$~(III), $-1.15$~(IV) and $-1.21$~(V). The inset shows magnetoresistance measurements for temperatures (top to bottom) $T$ = 0.671 K, 0.642 K, 0.610 K, 0.583 K, 0.551 K, and 0.522 K.}
\label{Fig-Theory-DWS}
\end{figure}

Both theoretical and experimental results show striking similarities which can be interpreted as follows. The phase boundary for a uniformly magnetized ferromagnetic film [curve  $\alpha=0$ and curve I in Fig. \ref{Fig-Theory-DWS}] is linear since there are no domains inside the ferromagnet and,  consequently, the effect of the magnetic stray field is negligible.

By applying a certain $H_{ret}<0$, negative domains are  introduced into the ferromagnetic film. For instance, for 
$H_{ret}/H_{c}=-1.05$ [curve II in Fig. \ref{Fig-Theory-DWS}(d)] the net magnetization is reduced to about 33\% of the saturation magnetization. The associated microscopic configuration is expected to be similar to the domain distribution shown in Fig. \ref{Fig-Magnetism}(e). By applying a negative external field, the stray field from the larger positive domains can be compensated  and superconductivity will nucleate \emph{locally} above these domains giving rise to a peak in the phase boundary located at negative fields. As a direct consequence of the increase of the ground energy of the ''particle-in-a-box'', the maximal critical temperature decreases drastically as the width of the positive domain decreases  [curves $\alpha=0.1$, $\alpha=0.2$ and curve II in Fig.~\ref{Fig-Theory-DWS}].

A further increase of $|H_{ret}|$ leads to a more pronounced
decrease of the positive domains, resulting in an even lower
$T_{c}$. This peak is now located at even higher negative fields, since the absolute value of the z-component of the field increases with decreasing domain size  [Fig.~\ref{Fig-Theory-DWS}(b)]. Simultaneously, the growth of negatively magnetized domains results in a more favorable OP nucleation above negative domains and, accordingly, a second peak in the critical temperature at $H>0$ develops [curve $\alpha=0.5$ and curve III in Fig.~\ref{Fig-Theory-DWS}]. The relative amplitude of these peaks is determined by the specific details of the magnetic domain structure. For a returning field of $H_{ret}/H_{c}=-1.15$ the remanent magnetization is close to zero, thus indicating the presence of an equal distribution of positive and negative domains. This domain structure gives rise to a nearly symmetric phase boundary [curve $\alpha=1$ and curve IV in Fig.~\ref{Fig-Theory-DWS}]. Similar phase boundaries with two maxima have already been observed in S/F hybrids, containing Nb and Pb films, and are attributed to domain wall superconductivity \cite{Yang-04, Gillijns-05, Yang-06}. For higher $H_{ret}$ values the first peak, located at negative fields, disappears, whereas the peak at positive fields shifts up in temperature and is displaced to a lower field [curve V in  Fig.~\ref{Fig-Theory-DWS}]. This second peak will eventually evolve in a linear phase boundary when the ferromagnetic film is fully magnetized in the negative direction.

It is worth mentioning that the good agreement between the experimental results and our simplified 1D model indicates that the used model captures the essential physics behind the magnetic confinement effect very well. Yet a small discrepancy exists near $H=0$ where theory predicts a small peak which is not observed in the experiment. This peak corresponds to a wide OP distribution, spreading over many periods of the magnetic field modulation~\cite{Aladyshkin-06}. We believe that this delocalized state is suppressed by the irregular shape of the domain distribution present in the real system.

We have applied the concept of tunable magnetic confinement also
to another S/F hybrid system, consisting of a superconducting Al
film covering an array of magnetic dots. The phase boundaries of
such a structure in different magnetic states are shown in
Fig.~\ref{Fig-FIS}. Due to the presence of the magnetic dots {\it
three} different areas, where the OP can be localized, are
present: above the positive or negative domains inside the
magnetic dot (similar to the bilayered system considered above)
and between the dots, where the local magnetic field is roughly
determined by the average magnetization of the dots. If the dots
are magnetized positively, there is a negative field between the
dots and vice versa. Note that there are no analogues of such
regions with zero magnetization in the S/F bilayers.

\begin{figure}[t!]
\caption{(color online) The phase boundaries $T_c(H)$ for an S/F
hybrid, consisting of an Al film and an array of magnetic dots, in the demagnetized (I), the completely magnetized (II) and several intermediate magnetic states:  $H_{ret}/H_{c}=-0.99$~(III), $-1.28$~(IV) and $-1.54$~(V), where $H_{c}=227~$mT.} \label{Fig-FIS}
\end{figure}

In the demagnetized state [curve I in Fig. \ref{Fig-FIS}] the
field in the region between the dots is approximately zero, accordingly superconductivity starts to nucleate at this position at relatively low magnetic fields. As a result, a linear phase
boundary centered at zero field is observed. At higher fields
($|H|\stackrel{_>}{_\sim}8$~mT) there is a clear deviation from
the expected linear behavior. Such magnetic bias can be explained
by the compensation of the magnetic field above the positive and
negative domains inside the magnetic dot, similar to the bilayered sample presented above. Interestingly this phase boundary combines field compensation effects in each of the three regions.

By magnetizing the dots positively (i) the amplitude of the field
in between the dots increases negatively and (ii) the typical size of the positive domains becomes larger than that for negative domains. As a result, the peak, associated with the OP
localization in between the dots, shifts towards positive fields (so called field-induced superconductivity) and a second local $T_c$ maximum, corresponding to the appearance of  superconductivity above the wider positive domains, appears,  while the OP  nucleation above narrower negative domains is suppressed [curves II-III in Fig. \ref{Fig-FIS}]. For negatively magnetized dots the reversed effect occurs [curves IV-V in Fig. \ref{Fig-FIS}]. It is important to note, that the amplitude of the main $T_c$ peak remains almost constant when the magnetic state of dots is  changed [compare the curves I-V in Fig.~\ref{Fig-FIS}(b)]. Indeed, since this peak corresponds to the nucleation of superconductivity in between the dots, the area of localization is almost independent of the dot's magnetic state.

Summarizing, we have studied tunable magnetic confinement of the
superconducting OP in different S/F hybrids, which originates from nonuniform magnetic fields induced by a ferromagnetic template. By manipulating the domain structure in the ferromagnet through appropriate demagnetizing procedures, one can drastically change the position of the confined OP and, as a
result, the shape of the phase boundary $T_c(H)$. In particular,
restricting the area, where optimal field compensation occurs, is
shown to induce a systematic reduction of the critical temperature of the superconducting transition. We have demonstrated that both domain-wall superconductivity and field-induced superconductivity are manifestations of the magnetic confinement effect in different magnetic structures.

This work was supported by the K.U.Leuven Research Fund
GOA/2004/02 program, the Belgian IUAP, the Fund for Scientific
Research -- Flanders (F.W.O.--Vlaanderen), the bilateral project
BIL/05/25 between Flanders and Russia, by the Russian Foundation
for Basic Research (A.Yu.A.) and by the F.W.O. fellowship
(A.V.S.).

\end{document}